\documentclass[a4 paper, 12 pt] {article}
\usepackage{graphics, graphicx}

\begin{document}
\title{\bf{Knotted Configurations with Arbitrary Hopf Index from the Eikonal
Equation}}
\author{A. Wereszczy\'{n}ski \thanks{wereszcz@alphas.if.uj.edu.pl}
       \\
       \\ Institute of Physics,  Jagiellonian University,
       \\ Reymonta 4, 30-059 Krak\'{o}w, Poland}
\maketitle
\begin{abstract}
The complex eikonal equation in $(3+1)$ dimensions is
investigated. It is shown that this equation generates many
multi-knot configurations with an arbitrary value of the Hopf
index. In general, these eikonal knots do not have the toroidal
symmetry. For example, a solution with topology of the trefoil
knot is found. Moreover, we show that the eikonal knots provide an
analytical framework in which qualitative (shape, topology) as
well as quantitative (energy) features of the Faddeev-Niemi
hopfions can be captured. It might suggest that the eikonal knots
can be helpful in construction of approximated (but analytical)
knotted solutions of the Faddeev-Skyrme-Niemi model.
\end{abstract}
%%%%%%%%%%%%%%%%%%%%%%%%%%%%%%%%%%%%%%%%%%%%%%%%%%%%%%%%%%%%%%%
\section{\bf{Introduction}}
%%%%%%%%%%%%%%%%%%%%%%%%%%%%%%%%%%%%%%%%%%%%%%%%%%%%%%%%%%%%%%%
Topological solitons i.e. stable, particle-like objects with a
non-vanishing topological charge occur in various contexts of the
theoretical physics. For example, as magnetic monopoles and
vortices seem to play crucial role in the problem of the
confinement of the quarks in the quantum chromodynamics \cite{wu},
\cite{thooft}. On the other hand, it is believed that other type
of solitons so-called cosmic strings and domain walls are
important for the time evolution of the universe and formation of
long range structures \cite{vilenkin}, \cite{davis}. Moreover, as
D-branes, they appear in the string theory as well. They are
observed also in various experiments in the condensed matter
physics (see for instance $^3He$ \cite{he3} and $^4He$ \cite{he4}
quantum liquids). In fact, richness of the possible application of
solitons is enormous.
\\
In particular, hopfions i.e. topological solitons with the
non-trivial Hopf index $Q_H \in \pi_3(S^2)$ have been recently
analyzed in a connection with the non-perturbative regime of the
gluodynamics. Namely, it has been suggested by Faddeev and Niemi
\cite{niemi1} that particles built only of the gauge field,
so-called glueballs, can be described as knotted solitons with a
non-vanishing value of the Hopf number. It is a natural extension
of the standard flux-tube picture of mesons where, due to the dual
Meissner effect, quark and anti-quark are confined by a tube of
the gauge field. In the case of absence of quark sources, such a
flux-tube should form a knotted, closed loop. Then, stability of
the configuration would be guaranteed by a non-zero value of the
topological charge. A model (so called the Faddeev-Skyrme-Niemi
model \cite{faddeev}), based on an effective classical three
component unit field (which is believed to represent all infrared
important degrees of freedom of the full quantum theory) has been
proposed \cite{niemi1}, \cite{cho}, \cite{shabanov}, \cite{kondo}.
In fact, using some numerical methods many topological solitons
with the Hopf index have been found \cite{hellmund},
\cite{battyde}, \cite{salo}. However, since all Fadd\-eev-Nie\-mi
knots are known only in the numerical form some crucial questions
(e.g. their stability) are still far from a satisfactory
understanding. The situation is even worse. For a particular value
of the Hopf index it is not proved which knot gives the stable
configuration. Because of the lack of the analytical solutions of
the Fadd\-eev-Skyr\-me-Nie\-mi model such problems as interaction
of hopfions, their scattering or formation of bound states have
not been solved yet (for some numerical results see \cite{ward}).
\\
On the other hand, in order to deal with hopfions in an analytical
way, many toy models have been constructed \cite{nicole},
\cite{aratyn}, \cite{ja1}, \cite{ja2}. In general, all of them are
invariant under the scaling transformation. It not only provides
the existence of hopfions but also gives an interesting way to
circumvent the Derrick theorem \cite{derick}. This idea is quite
old and has been originally proposed by Deser et al. \cite{deser}.
As a result, topological hopfions with arbitrary Hopf number have
been obtained. Unfortunately, on the contrary to the
Fadd\-eev-Nie\-mi knots, all toy hopfions possess toroidal
symmetry i.e. surfaces of the constant $n^3$ are toruses (they are
called 'unknots'). It strongly restricts applicability of these
models.
\\
The main aim of the present paper is to find a systematic way of
construction of analytical configurations with the Hopf charge,
which in general do not have the toroidal symmetry and can form
really knotted structures, as for instance the trefoil knot
observed in the Fadd\-eev-Skyr\-me-Nie\-mi effective model.
Moreover, our approach allows us to construct multi-knot
configurations, where knotted solutions are linked and form even
more complicated objects.
\\
Moreover, though obtained here knots do not satisfy the
Fadd\-eev-Skyr\-me-Nie\-mi equations of motion, there are
arguments which allow us to believe that our solutions (we call
them as eikonal knots) can have something to do with
Fadd\-eev-Nie\-mi hopfions. In our opinion this paper can be
regarded as a first step in construction of analytical topological
solutions in the Fadd\-eev-Skyr\-me-Nie\-mi model. The relation
between the eikonal knots and the Fadd\-eev-Nie\-mi hopfions will
be discussed more detailed in the last section.
\\
Our paper is organized as follows. In section 2 we test our method
taking into consideration the eikonal equation in the $(2+1)$
Minkowski space-time. In this case, obtained multi-soliton
configurations occur to be solutions of the well-known $O(3)$
sigma model. Thus, we are able to calculate the energy of the
solitons and analyze the Bogo\-mol\-ny inequality between energy
and the pertinent topological charge i.e. the winding number. We
show that all multi-soliton solutions saturate this inequality
regardless of the number and positions of the solitons. It must be
underlined that all results of this section are standard and very
well known. Nonetheless, we include this part to give a
pedagogical introduction to the next section.
\\
Section 3 is devoted to investigation of knotted solutions. Due to
that we solve the eikonal equation in the three dimensional space
and find in an analytical form multi-knot solutions with arbitrary
Hopf index. However, in this case no Lagrangian which possesses
all these configurations as solutions of the pertinent equations
of motion is known. Only one of our knots can be achieved in the
Nicole \cite{nicole} or Aratyn-Ferreira-Zimerman \cite{aratyn}
model.
\\
Finally, the connection between the eikonal knots and the
Faddeev-Niemi hopfions is discussed. We argue that our solutions
could give a reasonable approximation to the knotted solitons of
the Faddeev-Skyrme-Niemi model.
%%%%%%%%%%%%%%%%%%%%%%%%%%%%%%%%%%%%%%%%%%%%%%%%%%%%%%%%%%%%%%%
\section{\bf{$(2+1)$ dimensions: $O(3)$ sigma model}}
%%%%%%%%%%%%%%%%%%%%%%%%%%%%%%%%%%%%%%%%%%%%%%%%%%%%%%%%%%%%%%%
Let us start and introduce the basic equation of the pre\-sent
paper i.e. the complex eikonal equation
\begin{equation}
(\partial_{\nu} u)^2=0, \label{eikonal}
\end{equation}
in $(2+1)$ or $(3+1)$ dimensional space-time, where $u$ is a
complex scalar field. It is known that such a field can be
related, by means of the standard stereographic projection, with
an unit three component vector field $\vec{n} \in S^2$. Namely,
\begin{equation}
\vec{n}= \frac{1}{1+|u|^2} ( u+u^*, -i(u-u^*), |u|^2-1).
\label{stereograf}
\end{equation}
This vector field defines the topological contents of the model.
Depending on the number of the space dimensions and asymptotic
conditions this field can be treated as a map with $\pi_2(S^2)$ or
$\pi_3(S^2)$ topological charge.
\\
In this section we focus on the eikonal equation in $(2+1)$
dimensions. The main aim of this section is to consider how the
eikonal equation generates multi-soliton configurations. From our
point of view the two dimensional case can be regarded as a toy
model which should give us better understanding of the much more
complicated and physically interesting three dimensional case.
\\
In order to find solutions of equation (\ref{eikonal}), we
introduce the polar coordinates $r$ and $\phi$ and assume the
following Ansatz
\begin{equation}
u=\sum_{i=1}^N f_i(r) e^{ik_i \phi } + u_0, \label{anzatz O3}
\end{equation}
which is a generalization of the standard one-soliton An\-satz.
Here $N=1,2,3...$ and $k_i$ are integer numbers. Then $u$ is a
single valued function. Additionally $u_0$ is a complex constant.
After substituting it into the eikonal equation (\ref{eikonal})
one derives
\begin{equation}
\sum_{i=1}^N \left( f'^2_i -\frac{k^2_i}{r^2} f^2_i \right)
e^{2ik_i\phi} + 2 \sum_{j< i} \left( f'_j f'_i -\frac{k_jk_i}{r^2}
f_j f_i \right) e^{i(k_i+k_j)\phi} =0. \label{eqmot_O3_4}
\end{equation}
One can immediately check that it is solved by the following two
functions, parameterized by the positive integer numbers $k_j$
\begin{equation}
f_j=A_j r^{k_j} \label{sol_O3_1}
\end{equation}
and
\begin{equation}
f_j=B_j \frac{1}{r^{k_j}}. \label{sol_O3_2}
\end{equation}
Here $A_j$ and $B_j$ are arbitrary, in general complex, constants.
We express them in more useful polar form $A_j=a_je^{i\psi_i}$,
where $a_j, \psi_j$ are some real numbers. Thus, the solution
reads
\begin{equation}
u=\sum_{j=0}^N a_j r^{\pm k_j} e^{ik_j \phi }e^{i\psi_j},
\label{sol O3_u}
\end{equation}
where the constant $u_0=a_0e^{i\psi_0}$ with $n_0=0$ has been
included as well.
\\
From now, we restrict our investigation only to the family of the
solutions given by (\ref{sol_O3_1}). In the other words we have
derived the following configuration of the unit vector field
\begin{equation}
n^1= \frac{2\sum_{i=0}^N a_ir^{k_i} \cos (k_i \phi +\psi_i)
}{\sum_{i,j=0}^N r^{k_i+k_j} \cos [ (k_i-k_j)\phi
+(\psi_i-\psi_j)]+1} \label{soln_O3_1}
\end{equation}
\begin{equation}
n^2= \frac{2\sum_{i=0}^N a_ir^{k_i} \sin (k_i \phi +\psi_i)
}{\sum_{i,j=0}^N r^{k_i+k_j} \cos [(k_i-k_j)\phi+(\psi_i-\psi_j)]
+1} \label{soln_O3_2}
\end{equation}
\begin{equation}
n^3=\frac{\sum_{i,j=0}^N a_ia_jr^{k_i+k_j} \cos [(k_i-k_j)\phi
+(\psi_i-\psi_j)] -1}{\sum_{i,j=0}^N a_ia_jr^{k_i+k_j} \cos
[(k_i-k_j)\phi +(\psi_i-\psi_j)]+1}. \label{soln_O3_3}
\end{equation}
Let us shortly analyze above obtained solutions.
\\
First of all one could ask about the topological charge of the
solutions. The corresponding value of the winding number might be
calculated from the standard formula
\begin{equation}
Q=\frac{1}{8\pi} \int d^2x \epsilon^{ab} \vec{n} (\partial_a
\vec{n} \times \partial_b \vec{n}). \label{deg_n}
\end{equation}
However, since the introduced Ansatz is nothing else but a
particular (polynomial) rational map one can take into account a
well-known fact that the topological charge of any rational map of
the form $R(z)=p(z)$, where $p$ is a polynomial, is equal to
degree of this polynomial. Thus
\begin{equation}
Q=\mbox{max}\{ k_i, \;  i=1...N \}. \label{brawer}
\end{equation}
Quite interesting, we can notice that the total topological charge
of these solutions is fixed by the asymptotically leading term
i.e. by the biggest value of $k_i$ whereas the local distribution
of the topological solitons depends on all $k_i$ numbers.
\\
In fact, if we look at our solution at large $r$ then the vector
field wraps $Q$ times around the origin. As we discus it below,
our configuration appears to be a system of $L$ solitons with a
topological charge $Q_l, \; l=1..L$, where $L$ depends on $N, k_i$
and $a_0$. The total charge is a sum of $L$ individual charges
$$\sum_{l=1..L}Q_l=Q.$$
Let us now find the position of the solitons. It is defined as a
solution of the following condition
$$n^3=-1.$$
Thus points of location of the solitons fulfil the equation
\begin{equation}
\sum_{i,j=0}^N a_ia_jr^{k_i+k_j} \cos [(k_i-k_j)\phi
+(\psi_i-\psi_j)]=0. \label{position_solit_O3}
\end{equation}
Unfortunately, we are not able to find an exact solution of
equation (\ref{position_solit_O3}) for arbitrary $N$. Of course,
it can be easily done using some numerical methods. Let us
restrict our consideration to the two simplest but generic cases.
\\
We begin our analysis with $N=1$. This case, simply enough to find
exact solutions, admits various multi-so\-li\-ton configurations.
One can find that examples with more complicated Ansatz i.e.
larger $N$ seem not to differ drastically. The main features
remain unchanged.
\\
For $N=1$ equation (\ref{position_solit_O3}) takes the form
\begin{equation}
a_1^2 r^{2k_1}+2a_1a_0 \cos [k_1\phi +(\psi_1-\psi_0)]+a_0^2=0.
\label{position_solit_O3_N1}
\end{equation}
We see that there are $n_1$ solitons, each with unit topological
number, located symmetrically on the circle with radius
\begin{equation}
r=\left( \frac{a_1}{a_0} \right)^{\frac{1}{k_1}},
\label{position_solit_O3_sol_1r}
\end{equation}
in the points
\begin{equation}
 \phi= \frac{\pi -(\psi_1-\psi_0)+ 2l\pi}{k_1},
\label{position_solit_O3_sol_1}
\end{equation}
where $l=0,1...|k_1|-1$.
\\
Another simple but interesting example is the case with $N=2$ and
$a_0=0$. Then equation (\ref{position_solit_O3}) reads
\begin{equation}
\sum_{i,j=1}^2a_ia_jr^{k_i+k_j} \cos \left[ (k_i-k_j)\phi
+(\psi_i-\psi_j)\right]=0 \label{position_solit_O3_N2}
\end{equation}
The solitons are located in the following points
\begin{equation}
r=0,
\end{equation}
and
\begin{equation}
r=\left( \frac{a_1}{a_2} \right)^{\frac{1}{k_2-k_1}}, \; \; \; \;
\; \phi= \frac{\pi -(\psi_1-\psi_2)+ 2l\pi}{k_1 -k_2}
\label{position_solit_O3_sol}
\end{equation}
where $l=0,1...|k_1-k_2|-1$. It is clearly seen that there are two
different types of solitons. At the origin, we have a soliton with
the winding number equal to $\mbox{min} (k_1,k_2)$. Around it,
there are $|k_1-k_2|$ satellite solitons with an unit topological
charge.
\\
In Fig.1 - Fig.4 such soliton ensembles are demonstrated (we plot
$n^3$ component).
\begin{figure}
\center \resizebox{.4\textwidth}{!} {\includegraphics{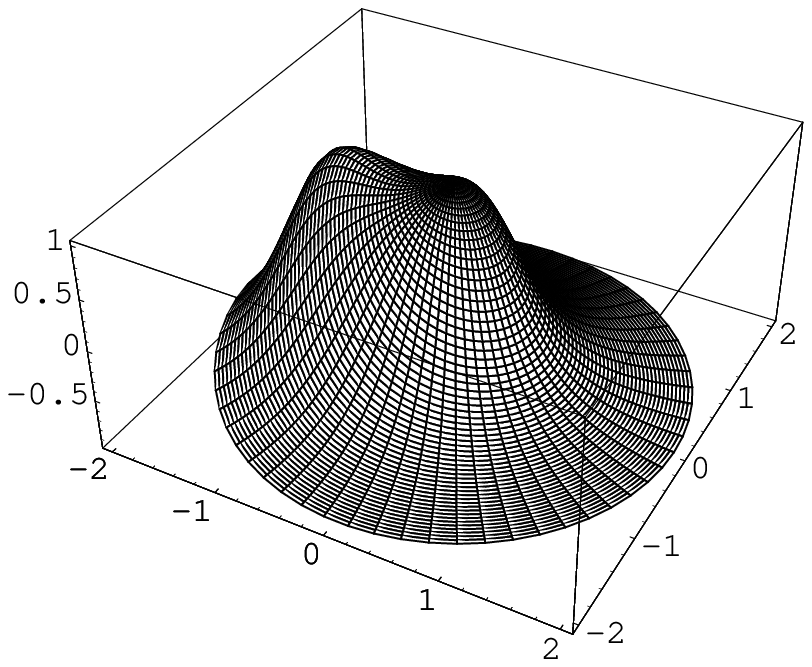}}
\caption{$k_1=1$ and $k_2=2$}

\center \resizebox{.4\textwidth}{!} {\includegraphics{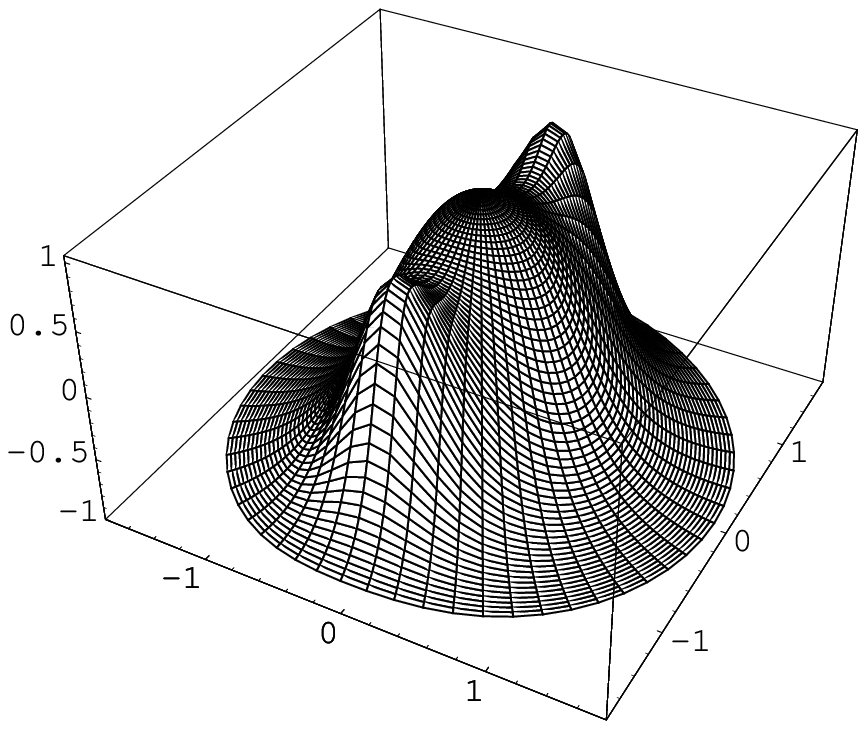}}
\caption{$k_1=1$ and $k_2=3$}
\end{figure}
\begin{figure}
\center \resizebox{.4\textwidth}{!} {\includegraphics{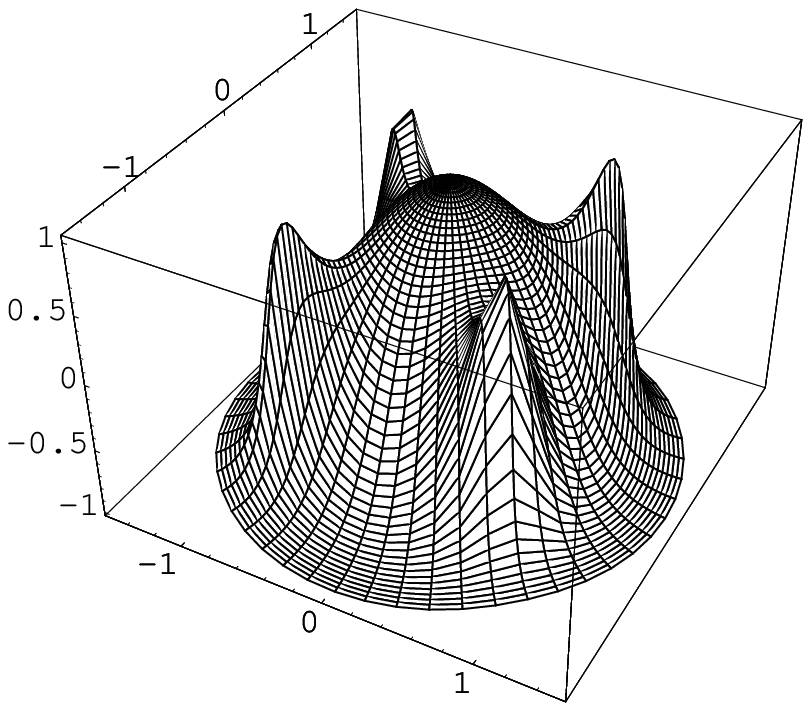}}
\caption{$k_1=1$ and $k_2=5$}

\center \resizebox{.4\textwidth}{!} {\includegraphics{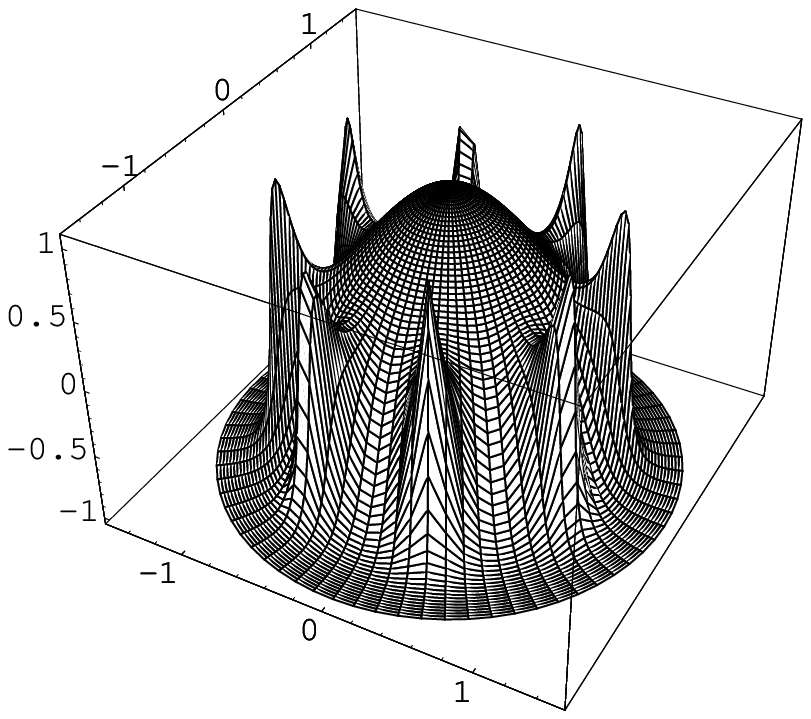}}
\caption{$k_1=1$ and $k_2=9$}
\end{figure}
\begin{figure}
\center \resizebox{.4\textwidth}{!} {\includegraphics{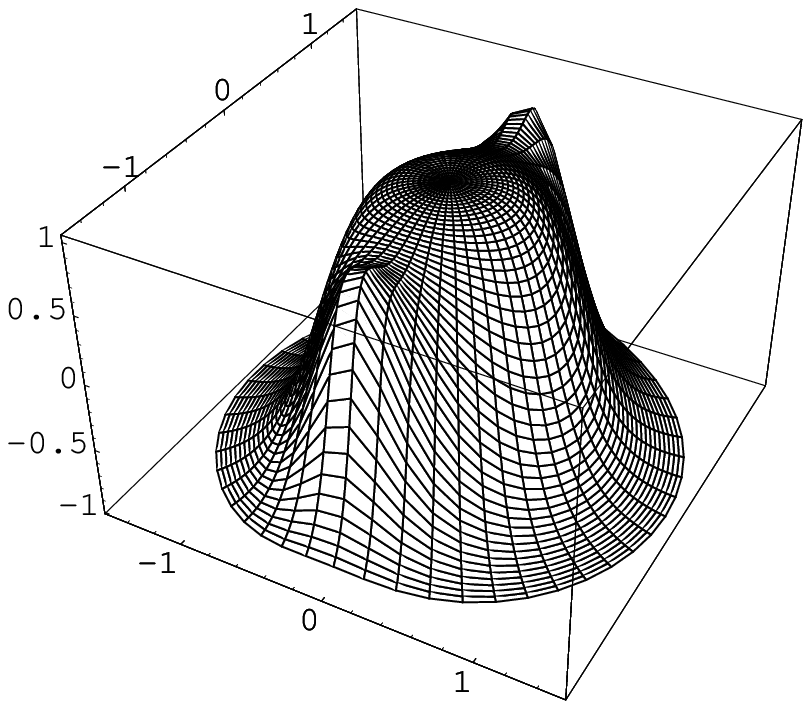}}
\caption{$k_1=2$ and $k_2=4$}
\end{figure}
For simplicity reason, we assume $\psi_1=\psi_2=0$ and
$a_1=a_2=1$. We see that there is a soliton with $Q=1$ at the
origin. Additionally, one, two, four or eight single satellite
solitons are shown. In Fig.5 the case with $k_1=2$ and $k_2=4$ is
plotted. It is very similar to Fig. 2 but now, the soliton located
at the origin is 'thicker' - it possesses $Q=2$ topological
charge. One can easily continue it and find more complicated
multi-soliton configurations.
\\ \\
It can be shown that presented configurations are also solutions
of a dynamical system i.e. the well-known non-linear $O(3)$ sigma
model
\begin{equation}
L=\frac{1}{2} (\partial_{\mu } \vec{n} )^2 \label{O(3)}.
\end{equation}
This fact permits us to call the solutions of the two dimensional
eikonal equation as solitons.
\\
In order to show it we take advantage of the stereographic
projection (\ref{stereograf}). Then the Lagrangian (\ref{O(3)})
takes the form
\begin{equation}
L= \frac{2 }{(1+|u|^2)^2} \partial_{\mu } u \partial^{\mu } u^*.
\label{O(3)_1}
\end{equation}
The pertinent equation of motion reads as follows
\begin{equation}
\frac{\partial_{\mu}\partial^{\mu} u}{(1+|u|^2)^2}
-2(\partial_{\nu} u)^2  \frac{u^*}{(1+|u|^2)^3} =0.
\label{eqmot_O3_1}
\end{equation}
It is straightforward to see that it is possible to introduce a
submodel defined by the following two equations: a dynamical
equation
\begin{equation}
\partial_{\mu}\partial^{\mu} u=0 \label{eqmot_O3_2}
\end{equation}
and a non-dynamical constrain
\begin{equation}
(\partial_{\nu} u)^2=0 \label{eqmot_O3_3},
\end{equation}
which is just the eikonal equation. One can notice that every
solution of the submodel fulfills the field equation for the
original model as well. However, it has to be underlined that the
space of solutions of the original model is much larger than the
restricted theory.
\\
Inserting the Ansatz (\ref{anzatz O3}) into the first formula
(\ref{eqmot_O3_2}) we obtain
\begin{equation}
\sum_{j=1}^N e^{ik_j\phi} \left( \frac{1}{r} \partial_r (r f'_j)
-\frac{k^2_j}{r^2} f_j \right) =0. \label{eqmot_O3-5}
\end{equation}
One can easily check that solutions (\ref{sol_O3_1}) and
(\ref{sol_O3_2}) satisfy this equation. It proves that our
multi-soliton configurations are not only solutions of the eikonal
equation but also are generated from Lagrangian (\ref{O(3)}).
Thus, we are able to calculate the corresponding energy. It is
easy to see that all solutions possess finite total energy. In
fact, $T_{00}$ part of the energy-momentum tensor
\begin{equation}
T_{00} = \frac{ 2 \sum_{i,j=1}^N a_ia_jk_ik_jr^{k_i+k_j-2} \cos
[(k_i-k_j)\phi +(\psi_i-\psi_j)]}{\left( \sum_{i,j=1}^N
a_ia_jr^{k_i+k_j} \cos [(k_i-k_j)\phi+(\psi_i-\psi_j)] +1
\right)^2} \label{t00}
\end{equation}
does not have any point-like singularities and tends to zero for
$r \rightarrow \infty$ sufficiently fast to assure finiteness of
the energy.
\\
We explicitly calculate the total energy in the case $N=2$. Using
previously obtained solution we find that
\begin{equation}
E=2 \int_0^{\infty} \int_0^{2\pi} \frac{r dr d\phi}{r^2}
 \frac{k_1^2
r^{2k_1}+k_2^2 r^{2k_2}+2k_1k_2r^{k_1+k_2}\cos (k_1-k_2)
\phi}{(1+r^{2k_1}+r^{2k_2}+2r^{k_1+k_2}\cos (k_1-k_2) \phi)^2}.
\label{energy1}
\end{equation}
This integral can be evaluated and one obtains
\begin{equation}
E=4\pi \mbox{max} (k_1, k_2). \label{energy2}
\end{equation}
It is equivalent to the following relation
\begin{equation}
E=4\pi Q. \label{en_q_rel}
\end{equation}
The multi-soliton solutions saturate the famous energy-char\-ge
inequality for $O(3)$ sigma model i.e. $E \geq 4\pi |Q|$. It means
that they are stable. It is not possible to have less energy
solutions with a fixed value of the total topological number. It
is worth to stress that a single soliton solution with $n$
topological index has exactly the same energy as collection of $n$
solitons with an unit charge. Moreover, the energy of the
multi-soliton solutions does not depend on the relative position
of the solitons. It gives us the possibility to analyze scattering
of the solitons using the standard moduli-space method.
\\
As it was said before, all result presented in this section are
well known. However, we have reproduce them from a new point of
view i.e. using the complex eikonal equation. It is nothing
surprising if we observe that the eikonal equation in the two
dimensions leads to a generalization of the Cauchy-Riemann
equations
\begin{equation}
u_z u_{\bar{z}}=0, \label{cr eik}
\end{equation}
where $z=x+iy$. Since all Baby Skyrmions are rational functions of
the $z$ variable (or $\bar{z}$) thus they can be in the natural
way found in the eikonal equation.
%%%%%%%%%%%%%%%%%%%%%%%%%%%%%%%%%%%%%%%%%%%%%%%%%%%%%%%%%%%%%%%
\section{\bf{$(3+1)$ dimensions: the eikonal knots}}
%%%%%%%%%%%%%%%%%%%%%%%%%%%%%%%%%%%%%%%%%%%%%%%%%%%%%%%%%%%%%%%
Let us now turn to the complex scalar field $u$ living in the
$3+1$ dimensional Minkowski space-time. Analogously as in the
previous section such a complex field can be used, via the
stereographic projection, to parameterize a three component unit
vector field $\vec{n}$:
\begin{equation}
\vec{n}= \frac{1}{1+|u|^2} ( u+u^*, -i(u-u^*), |u|^2-1).
\label{stereograf1}
\end{equation}
Due to the fact that all static configurations, such that $\vec{n}
\rightarrow \vec{n}_0=\overrightarrow{\mbox{const.}}$ for
$|\vec{x}| \rightarrow \infty$, are maps $\vec{n} : R^3 \cup
\{\infty \} \rightarrow S^2$, can be divided into disconnected
classes and characterized by a pertinent topological charge,
so-called Hopf index $Q_H \in \pi_3(S^2)$. In this section we will
show how such configurations can be generated by means of the
complex eikonal equation in the three space dimensions
\footnote{Appearance of knots as solutions of the complex eikonal
equation has been originally observed by Adam \cite{adam}}
\begin{equation}
\partial_i u \partial^i u =0. \label{eikonal1}
\end{equation}
In order to find exact solutions we assume the toroidal symmetry
of the problem and introduce the toroidal coordinates
$$ x=\frac{\tilde{a}}{q} \sinh \eta \cos \phi , $$
$$ y=\frac{\tilde{a}}{q} \sinh \eta \sin \phi , $$
\begin{equation}
z=\frac{\tilde{a}}{q} \sin \xi ,\label{tor_coord}
\end{equation}
where $q=\cosh \eta -\cos \xi $ and $\tilde{a}>0$ is a constant of
dimension of length fixing the scale. Moreover, we propose a
generalized version of the Aratyn-Ferreira-Zimerman-Adam Ansatz
\cite{aratyn}, \cite{adam} given by the following formula
\begin{equation}
u= \sum_{j=1}^N f_j(\eta ) e^{i(m_j \xi + k_j \phi)}+c,
\label{anzatz}
\end{equation}
where $m_i, \; k_i$ are integer numbers whereas $f_i$ are unknown
real functions depending only on the $\eta $ coordinate.
Additionally, $c$ is a complex number. Inserting our Ansatz
(\ref{anzatz}) into the eikonal equation (\ref{eikonal1}) one
derives
$$
0=\sum_{j=1}^N e^{2i(m_j \xi + k_j \phi) } \left( f'^2_j -
\left(m_j^2 +\frac{k^2_j}{\sinh^2 \eta}\right) f_j^2 \right) +$$
\begin{equation}
2\sum_{j<l} e^{i((m_j+m_l) \xi + (k_j+k_l) \phi) }  \left(
f'_jf'_l -\left(m_jm_l +\frac{k_jk_l}{\sinh^2 \eta} \right) f_jf_l
\right). \label{eikonal2}
\end{equation}
Thus the unknown shape functions $f_i$ should obey the following
equations
\begin{equation}
 f'^2_j - \left(m_j^2 +\frac{k^2_j}{\sinh^2 \eta}\right)
f_j^2 =0, \label{cond1}
\end{equation}
for $ j=1...N$ and
\begin{equation}
f'_jf'_l -\left(m_jm_l +\frac{k_jk_l}{\sinh^2 \eta} \right) f_jf_l
=0,  \label{cond2}
\end{equation}
for all $j\neq l$. The first set of the equations can be rewritten
into the form
\begin{equation}
f'_j=\pm \sqrt{\left(m_j^2 +\frac{k_j^2}{\sinh^2 \eta} \right)}
f_j. \label{sol_cond1}
\end{equation}
In the case of the positive sing we obtain solutions which have
been originally found by Adam \cite{adam}
\begin{equation}
f_j=A_j \sinh^{|k_j|} \eta \frac{\left( |m_j| \cosh \eta +
\sqrt{k_j^2 +m_j^2\sinh^2 \eta } \right)^{|m_j|}}{\left( |k_j|
\cosh \eta + \sqrt{k_j^2 +m_j^2\sinh^2 \eta } \right)^{|k_j|}}.
\label{sol_plus}
\end{equation}
They correspond to the following asymptotic value of the unit
vector field
\begin{equation}
\vec{n} \rightarrow \frac{1}{1+|c|^2} (c+c^*,-i(c-c^*),|c|^2-1) \;
\; \; \mbox{as} \; \; \eta \rightarrow 0 \label{bound1}
\end{equation}
and
\begin{equation}
\vec{n} \rightarrow (0,0,1) \; \; \; \mbox{as} \; \; \eta
\rightarrow \infty. \label{bound2}
\end{equation}
For the minus sing the solutions read
\begin{equation}
f_j=\frac{B_j}{\sinh^{|k_j|} \eta} \frac{\left( |k_j| \cosh \eta +
\sqrt{k_j^2 +m_j^2\sinh^2 \eta } \right)^{|k_j|}}{\left( |m_j|
\cosh \eta + \sqrt{k_j^2 +m_j^2\sinh^2 \eta } \right)^{|m_j|}},
\label{sol_minus}
\end{equation}
and the asymptotic behavior of the unit field is
\begin{equation}
\vec{n} \rightarrow (0,0,1) \; \; \; \mbox{as} \; \; \eta
\rightarrow 0 \label{bound3}
\end{equation}
and
\begin{equation}
\vec{n} \rightarrow \frac{1}{1+|c|^2} (c+c^*,-i(c-c^*),|c|^2-1) \;
\; \; \mbox{as} \; \; \eta \rightarrow \infty. \label{bound4}
\end{equation}
Let us now consider the second set of equations (\ref{cond2}) and
express constants as before i.e. $A_j=a_je^{i\psi_j}$. We also
take advantage of the fact that every $f_i$ has to fulfill
equation (\ref{cond1}). Then, inserting (\ref{sol_cond1}) into
(\ref{cond2}) we obtain that
\begin{equation}
\sqrt{\left(m_j^2 +\frac{k_j^2}{\sinh^2 \eta} \right)}
\sqrt{\left(m_l^2 +\frac{k_l^2}{\sinh^2 \eta} \right)}= m_jm_l
+\frac{k_jk_l}{\sinh^2 \eta}. \label{sol1_cond2}
\end{equation}
It leads to the relation
\begin{equation}
m_j^2 k_l^2+m_l^2k_j^2=2m_jm_lk_jk_l \label{sol2_cond2}
\end{equation}
or
\begin{equation}
(m_jk_l - m_lk_j)^2=0. \label{sol3_cond2}
\end{equation}
Finally, we derive the consistency conditions relating the integer
constants included in Ansatz (\ref{anzatz})
\begin{equation}
\frac{m_j}{k_j}=\frac{m_l}{k_l}, \; \; \; j,l=1...N
\label{sol_cond2}
\end{equation}
In other words, our Ansatz (\ref{anzatz}) is a solution of the
eikonal equation (with functions $f$ given by (\ref{sol_plus}) or
(\ref{sol_minus})) only if the ratio between the parameters $k_i$
and $m_i$ is a constant number.
\\
It is easy to notice that one can find more general solution of
the complex eikonal equation than the Ansatz. In fact, using the
simplest one-component solution with $m=k=1$
\begin{equation}
u_0= \frac{1}{\sinh \eta} e^{i(\xi +\phi)}, \label{simplest}
\end{equation}
we are able to generate other solutions. It follows from the
observation that any function of this solution solves the eikonal
equation as well. Thus
\begin{equation}
u=F(u_0), \label{general sol}
\end{equation}
where $F$ is any reasonable function, gives a new solution
\cite{adam}. Now, our Ansatz can be derived by acting a polynomial
function $F$ on the fundamental solution $u_0$ i.e.
$F(u_0)=c_o+c_1u+...+c_N u^N$.
\\
Calculation of the total Hopf index corresponding to the upper
obtained solutions can be carried out analogously as in the
two-dimensional case. Then, one obtains that
\begin{equation}
Q_H=-\mbox{max} \{k_im_i, i=1...N\}. \label{hopf_charge}
\end{equation}
Of course, it can be also found directly by calculation how many
times the vector field $\vec{n}$ wraps in the angular directions.
For $N=1$ and $c_0=0$ we see that $\vec{n}$ wraps $m$ times around
$\xi$-direction and $k$ times around $\phi$-direction, giving
$Q_H=-mk$. One can also see that for a non-vanishing $c$ (but
still for the one-knot configuration) the vector field behaves
identically. Thus, the topological charge is still $Q_H=-mk$. For
the multi-knot case, one has to not only add topological charges
of the elementary hopfions but also take into account the linking
number. As we discuss it in the next subsection, it leads to the
same total Hopf index i.e. $Q_H=-mk$. Analogous calculations can
be carried in the case of $N=2$.
\\
It is worth to stress that analogously to the winding number in
the $O(3)$ sigma model in $(2+1)$ dimensions, the total Hopf index
is fixed by the asymptotically leading term in our Ansatz. The
other components of the Ansatz affect only the local topological
structure of the solution.
\\
The position of the solution can be easily found if we recall that
in the core of a knot the vector field takes the opposite value
than at the spatial infinity
\begin{equation}
\vec{n}_0=-\vec{n}^{\infty},  \label{def pos1}
\end{equation}
where
\begin{equation}
\vec{n}^{\infty}=\lim_{\vec{x} \rightarrow \infty} \vec{n}=
\lim_{\eta \rightarrow 0} \vec{n}. \label{def pos2}
\end{equation}
Then a knotted solution is represented by a curve corresponding to
$\vec{n}=\vec{n}_0$.
\\
In further considerations we take solution (\ref{sol_minus}), for
which $f\rightarrow \infty$ as $\eta \rightarrow 0$ and
$\vec{n}^{\infty}=(0,0,1)$. Thus the core of a knot is located at
$\vec{n}_0=(0,0,-1)$. Therefore, it is given by a curve being a
solution to the equation
$$
\sum_{i,j=1}^N a_ia_j g_{i}g_{j} \cos[ (m_i-m_j)\xi +(k_i-k_j)\phi
+(\psi_i-\psi_j)] +$$
\begin{equation}
2c_0\sum_{i=1}^N a_ig_i \cos[m_i \xi +k_i \phi +
(\psi_1-\psi_2)+\alpha_0]+c_0^2=0, \label{position_solit_gen}
\end{equation}
where $f_i(\eta)=A_ig_i(\eta)$ and $c=c_0e^{i\alpha_0}$. The
simplest but sufficiently interesting cases with $N=1$ as well as
$N=2$ are analyzed below.
%%%%%%%%%%%%%%%%%%%%%%%%%%%%%%%%%%%%%%%%%%%%%%%%%%%%
\subsection{N=1 case}
%%%%%%%%%%%%%%%%%%%%%%%%%%%%%%%%%%%%%%%%%%%%%%%%%%%%
For $N=1$ the last equation can be rewritten in the following form
\begin{equation}
a^2g^2+2c_0 ag \cos[m \xi +k \phi + \psi +\alpha_0]+c_0^2=0.
\label{position_solit_gen_1}
\end{equation}
Thus, knots are located at
\begin{equation}
g(\eta_0)=\frac{c_0}{a}, \;  \; \; m \xi +k \phi =\pi- \psi
-\alpha_0 +2\pi l, \; \; \; l=0,1...L-1. \label{pos_sol_curve1}
\end{equation}
\begin{figure}
\center \resizebox{.4\textwidth}{!}
{\includegraphics{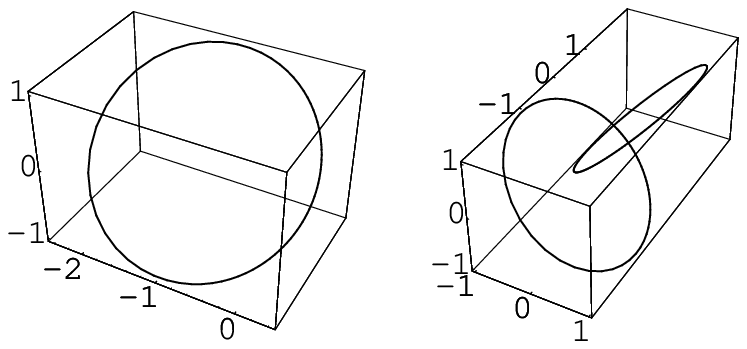}} \caption{$m=k=1$ and $m=k=2$}

\center \resizebox{.4\textwidth}{!}
{\includegraphics{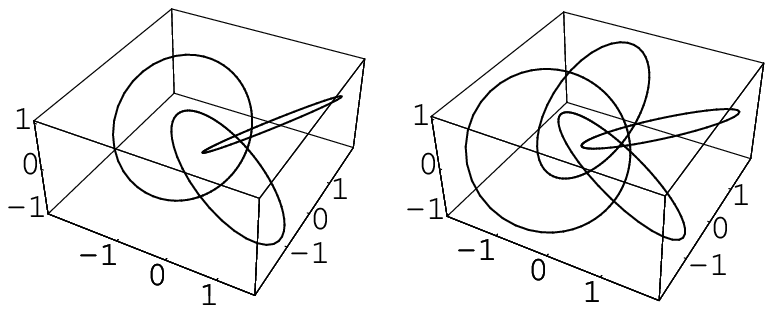}} \caption{$m=k=3$ and $m=k=4$}
\end{figure}
Due to the fact that $g$ is the monotonic function from $\infty$
to $0$, there is exactly one $\eta_0$ satisfies upper condition.
Thus, obtained configuration is given by a closed curve (or
curves) (\ref{pos_sol_curve1}) wrapped on a torus $\eta=\eta_0$.
In general, for fixed values of the parameters $m,k$ one finds
that the number $L$ of the elementary knots is equal to the
greatest common divisor $m$ and $k$. The whole multi-knot solution
is a collection of such elementary knots which are linked
together. Of course, every elementary knot should be treated in
the same manner as others so they all carry the identical
topological charge $Q_e= -pq$, where $p,q$ are relative prime
numbers and $\frac{m}{k}=\frac{p}{q}$. We immediately see that
simple sum of all charges of elementary knots is not equal to the
total topological number $Q=-km$. In order to correctly calculate
the topological charge one has to take into account the linking
number $N_L$ between elementary knots as well. Finally, we derive
that
\begin{equation}
Q_H=K \cdot Q_e - N_L. \label{linking}
\end{equation}
Correctness of this formula will be checked in some (but
sufficiently general) cases.
\begin{figure}
\center \resizebox{.4\textwidth}{!}
{\includegraphics{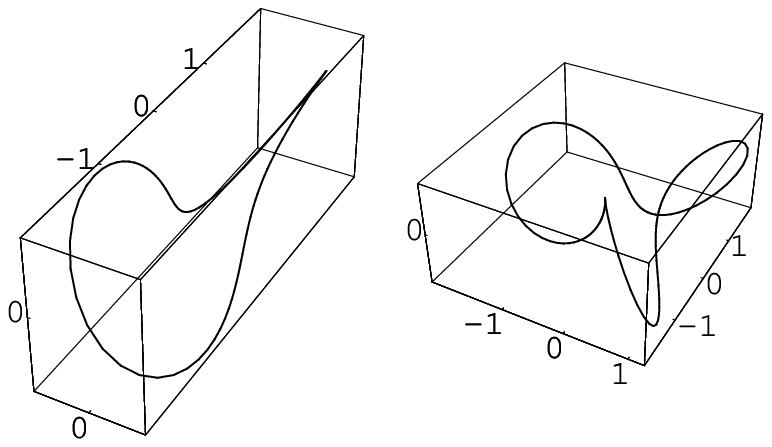}} \caption{$m=1,k=2$ and
$m=1,k=3$}

\center \resizebox{.4\textwidth}{!}
{\includegraphics{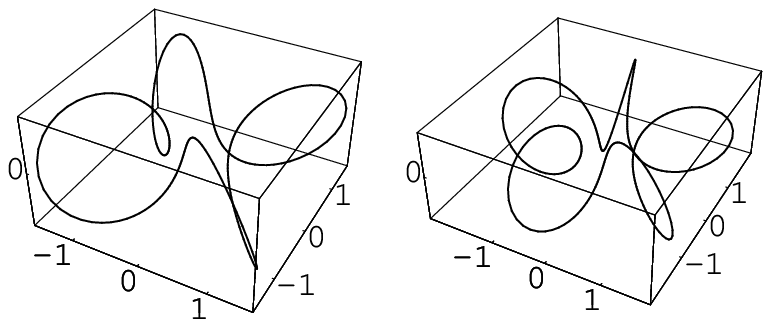}} \caption{$m=1,k=4$ and
$m=1,k=5$}

\center \resizebox{.4\textwidth}{!}
{\includegraphics{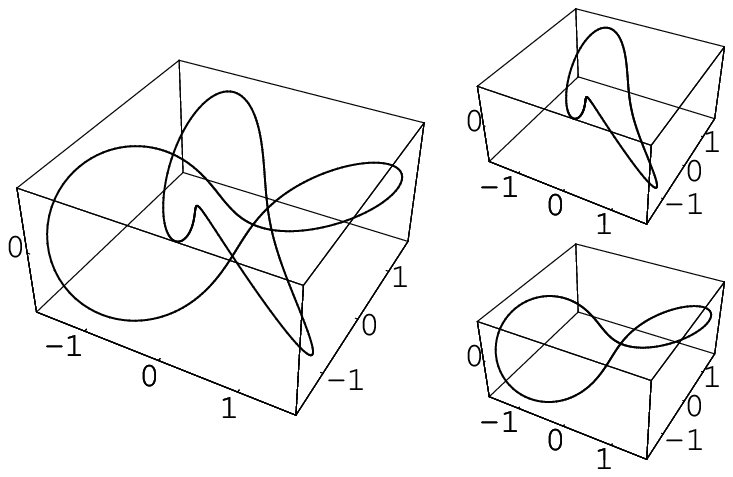}} \caption{$m=2,k=4$ with
elementary knots}
\end{figure}
Let us discus some of the obtained multi-knot configurations in
details. For simplicity we assume $\psi_1=\psi_2=0$, $a_1=a_2=1$.
In Fig. 6 the simplest eikonal knot with $m=k=1$ is presented. The
position of the knot is given by a circle and this configuration
possesses toroidal symmetry. In the same figure the case with
$m=k=2$ is demonstrated as well. As one could expect such a
configuration consists of two elementary knots with $Q_e=-1$
(circles) which are linked together. The linking number is $N_L=2$
so this configuration has the total charge $Q=-4$, what is in
accordance with the formula (\ref{linking}). Other multi-knot
configurations built of the elementary knots with $Q_e=-1$ are
plotted in Fig. 7 ($m=k=3$ and $m=k=4$). One can easily check that
(\ref{linking}) is fulfilled as well.
\\
The more sophisticated case is shown in Fig. 8, where a single
knot solution for $m=1, k=2$ as well as for $m=1, k=3$ is
presented. In contradistinction to the previously discussed knot
this configuration does not have the toroidal symmetry (of course,
one can easily restore the toroidal symmetry by setting $c=0$). In
Fig. 9 further examples with $m=1, k=4$ and $m=1, k=5$ are shown.
Now, it is obvious how this type of knots (i.e. with $R \equiv
\frac{k}{m} \in \mathcal{N}$) looks like. They cross $2R$ times
the $xy$-plane, or in the other words they wrap $R$ times
'vertically' on the torus $\eta_0$.
\begin{figure}

\center \resizebox{.4\textwidth}{!}
{\includegraphics{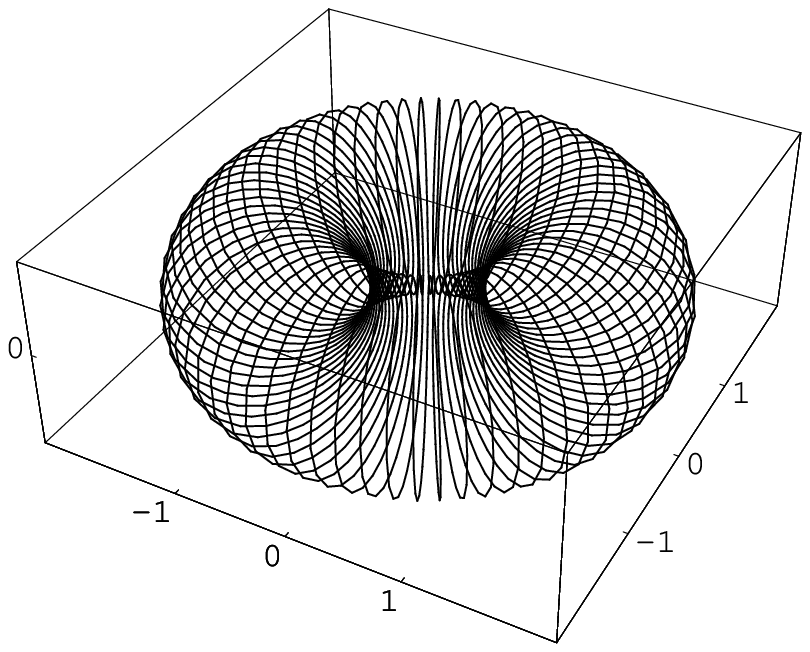}} \caption{$m=1,k=70$}

\center \resizebox{.4\textwidth}{!}
{\includegraphics{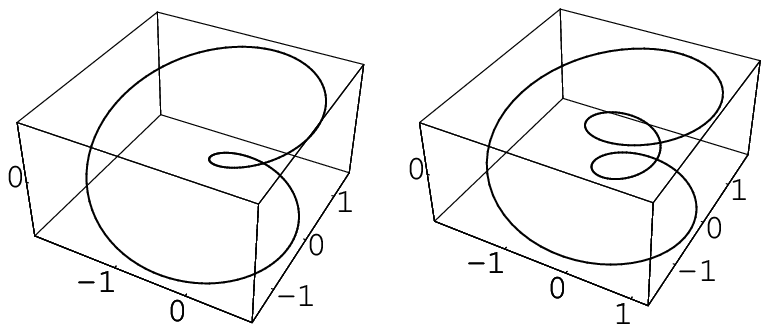}} \caption{$m=2,k=1$ and
$m=3,k=1$}
\end{figure}
\begin{figure}
\center \resizebox{.4\textwidth}{!}
{\includegraphics{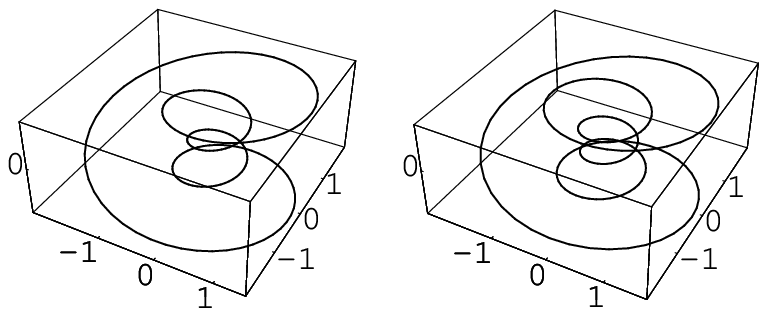}} \caption{$m=4,k=1$ and
$m=5,k=1$}

\center \resizebox{.4\textwidth}{!}
{\includegraphics{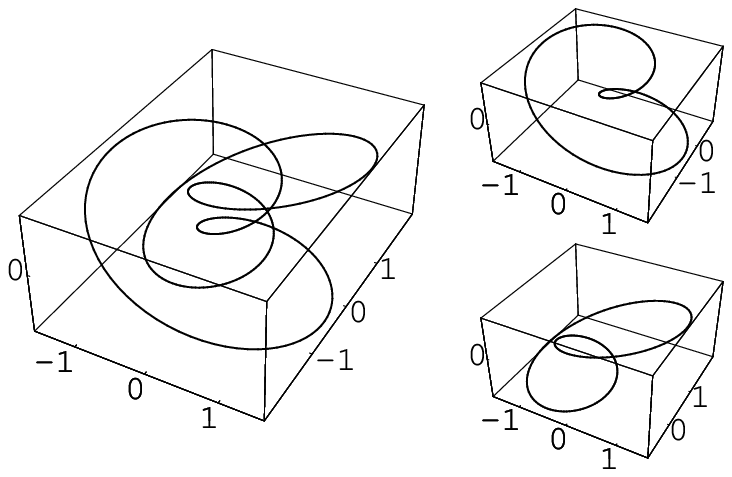}} \caption{$m=4,k=2$ with
elementary knots}

\center \resizebox{.4\textwidth}{!}
{\includegraphics{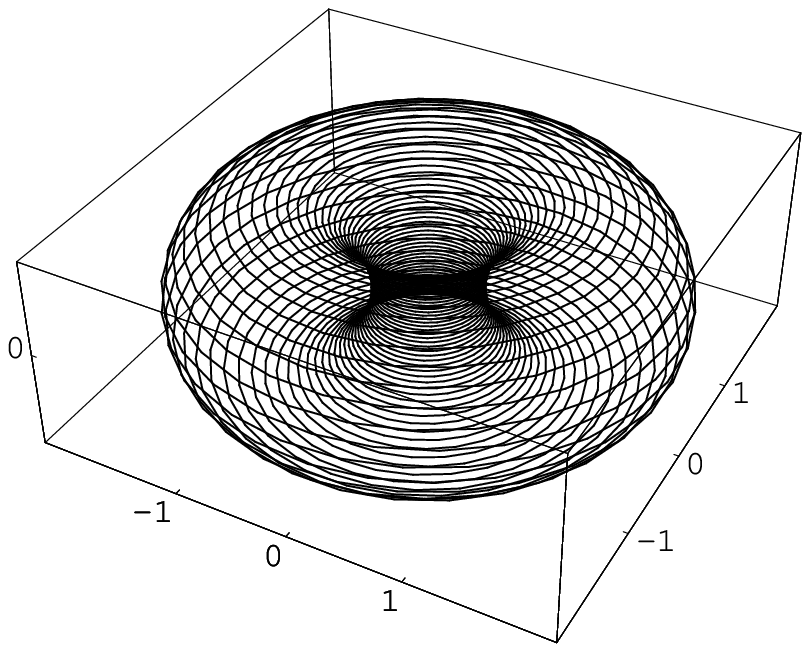}} \caption{$m=70,k=1$}
\end{figure}
\begin{figure}
\center \resizebox{.4\textwidth}{!}
{\includegraphics{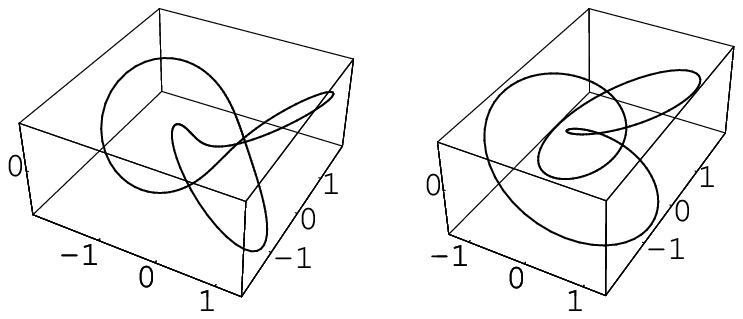}} \caption{$m=2,k=3$ and
$m=3,k=2$}

\center \resizebox{.4\textwidth}{!}
{\includegraphics{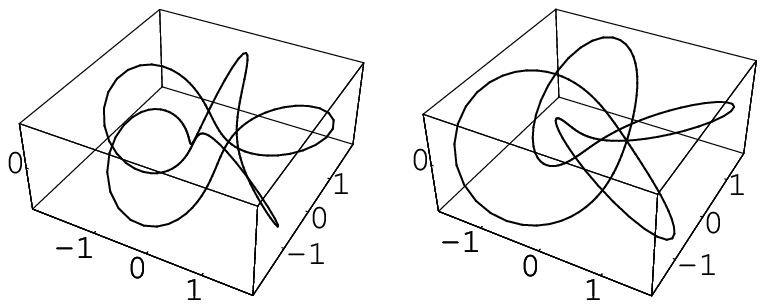}} \caption{$m=2,k=5$ and
$m=3,k=4$}

\center \resizebox{.4\textwidth}{!}
{\includegraphics{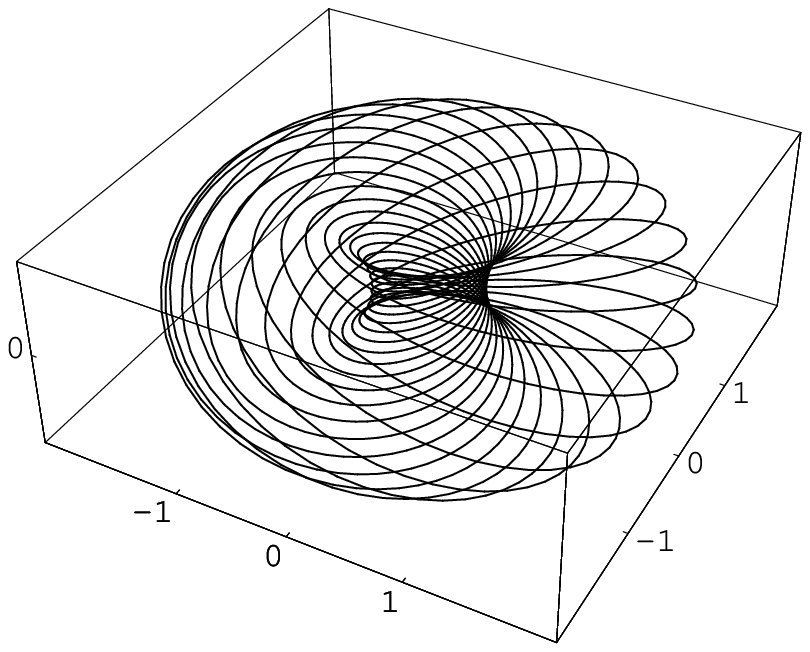}} \caption{$m=30,k=17$}
\end{figure}
Another configuration with two elementary knots is presented in
Fig. 10. Also in this case relation (\ref{linking}) is satisfied -
the corresponding linking number is $N_L=4$ and $Q_e=-2$. In Fig.
11 a knot with $m=1, k=70$ is plotted. It is clearly visible that
the knot is situated on a torus. Another simple type of solutions
can be obtained for $\frac{1}{R} \in N$. Such knots with $m=2,
k=1$ and $m=3, k=1$ are shown in Fig. 12. Here, the knot wraps $R$
times but in the 'horizontal' direction on the torus i.e. around
$z$-axis. Knots with higher topological charges ($m=4, k=1$ and
$m=5, k=1$) are plotted in Fig. 13. In Fig. 14 the simplest
two-knot configuration of this type is demonstrated whereas in Fig
15 one can find a solution with $m=70, n=1$. It is straightforward
to notice that, in spite of the fact that upper discussed knots
belong to distinct topological classes, the curves describing
their position are topological equivalent to a simple circle.
\\
More complicated and really knotted configurations have been found
for $R=\frac{p}{q}$, where $p,q $ are relative prime numbers
distinct from one. The simplest trefoil knots with $R=3/2$ and
$R=2/3$ are presented in Fig. 16. In both cases the Hopf charge is
$Q_H=-6$. In Fig. 17 farther examples of knots with $R=2/5$ as
well as $R=3/4$ are plotted. In Fig. 18 a really highly knotted
solution with $Q_H=-510$ is presented. We see that any Hopf
solution with $R=\frac{p}{q}$ wraps simultaneously $q$ times
around $z$-axis and $p$ times around circle $\eta=\infty$.
%%%%%%%%%%%%%%%%%%%%%%%%%%%%%%%%%%%%%%%%%%%%%%%%%%%%
\subsection{N=2 case}
%%%%%%%%%%%%%%%%%%%%%%%%%%%%%%%%%%%%%%%%%%%%%%%%%%%%
Let us now consider the second simple case i.e. we take $N=2$ and
put $c=0$. It enables us to construct a new class of multi-knot
configurations which differ from previously described. Equation
(\ref{position_solit_gen}) takes the form
\begin{equation}
0= a_1^2g_{1}^2 + a_2^2g_{2}^2+ 2 a_1a_2g_{1}g_{2} \cos[
(m_1-m_2)\xi +(k_1-k_2)\phi +(\psi_1-\psi_2)] .
\label{position_solit}
\end{equation}
One can solve it and obtain the position of knots. In this case we
can distinguish two sorts of solutions. Namely, a central knot
located at
$$
\eta=\infty$$
\begin{figure}
\center \resizebox{.4\textwidth}{!}
{\includegraphics{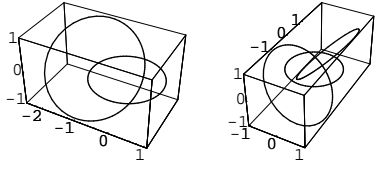}} \caption{$m_1=k_1=1$, $m_2=k_2=2$
and $m_1=k_1=1$, $m_2=k_2=3$}
\end{figure}
and satellite knots
\begin{equation}
\frac{g_{1}}{g_{2}}=\frac{a_2}{a_2}, \; \; \; \; \; (m_1-m_2) \xi
+ (k_1-k_2) \phi= \pi -(\psi_1-\psi_2)+ 2l\pi,
\label{position_solit_sol}
\end{equation}
where $l=0,...|L_1-L_2|-1$. It should be notice that the condition
$\frac{g_1}{g_2}=\frac{a_2}{a_2} $ can be always satisfied. It is
due to the fact that $g_1/g_2$ is a function smoothly and
monotonically interpolating between $\infty$ and $0$ (of course if
$k_1 \neq k_2)$. From (\ref{position_solit_sol}) we find that for
a fixed value of the parameters $m_i,k_i$ there are $|L_1-L_2|$
knots: one in $\eta = \infty$ (circle) and $|L_1-L_2|-1$ satellite
knots (loops) which can take various, quite complicated and
topologically inequivalent shapes. Additionally, one can observe
that the Hopf index of the central knot is $Q_c=-\mbox{min}
\{k_im_i, i=1..N \}$.
\begin{figure}
\center \resizebox{.4\textwidth}{!}
{\includegraphics{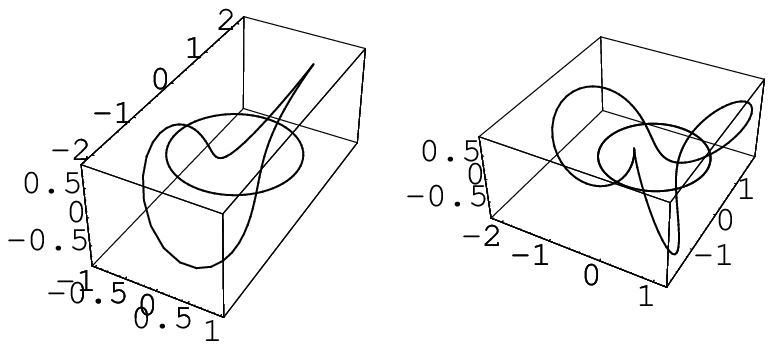}} \caption{$m_1=2, k_1=4$, $m_2=1,
k_2=2;$ $m_1=2, k_1=6$, $m_2=1, k_2=3$}

\center \resizebox{.4\textwidth}{!}
{\includegraphics{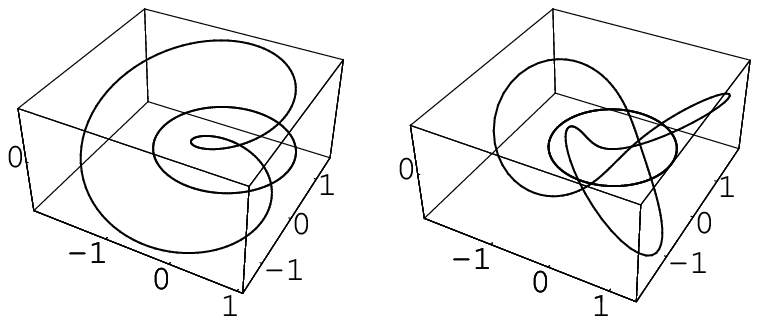}} \caption{$m_1=4, k_1=2$, $m_2=2,
k_2=1;$ $m_1=4, k_1=6$, $m_2=2, k_2=3$}
\end{figure}
In Fig. 19 the simplest types of solutions, with $m_1=k_1=1$,
$m_2=k_2=2$ and $m_1=k_1=1$, $m_2=k_2=3$ are demonstrated. We see
that they are very similar to the corresponding solutions with
$N=1$ (see Fig. 6 and Fig. 7). More complicated situations are
plotted in Fig. 20 and Fig. 21. In all configurations the central
knot is clearly visible as a circle around origin, whereas knots
known form the previous subsection wrap around this central knot.
\\ \\
It is obvious to notice that presented solutions (even one-knot
configurations) do not possess toroidal symmetry. Surfaces of a
constant value of the third component $n^3$ are not (in general)
toruses. In this manner they differ profoundly from the standard
knotted soliton configurations previously presented in the
literature \cite{nicole}, \cite{aratyn}, \cite{ja2}. Here, the
position of a knot depends on a (constant) value of the radial
coordinate $\eta $ as well as on the angular coordinates $\xi, \,
\phi$. Up to our best knowledge non-toroidal knots have been not,
in the exact form, presented in the literature yet.
\\
One has to be aware that all knots found in this paper are
solutions of the complex eikonal equation only. It is not known
any Lagrangian which would give these multi-knot configurations as
solutions of the pertinent equations of motion. They differ from
toroidal solitons obtained recently in the
Aratyn-Ferreira-Zimerman (and its generalizations) and the Nicole
model. However, it is worth to notice that the simplest one knot
state with $k=m=1$
\begin{equation}
u(\eta,\xi ,\phi)=\frac{1}{\sinh \eta} e^{i(\xi+\phi)}
\label{sol_Q1}
\end{equation}
is identical to the soliton with $Q_H=-1$ obtained in these models
\cite{nicole}, \cite{aratyn}. As we do not know the form of the
Lagrangian we are not able to calculate the energy corresponding
to the obtained multi-knot solutions. Thus, their stability and
saturation of the Vakulenko-Kapitansky inequality \cite{vakulenko}
are still open problems.
%%%%%%%%%%%%%%%%%%%%%%%%%%%%%%%%%%%%%%%%%%%%%%%%%%%%%%%%%%%%%%%
\section{\bf{The eikonal knots and
the Faddeev-Nie\-mi hopfions}}
%%%%%%%%%%%%%%%%%%%%%%%%%%%%%%%%%%%%%%%%%%%%%%%%%%%%%%%%%%%%%%%
In spite of the upper mentioned problems our multi-knot
configurations become more physically interesting, and potentially
can have realistic applications, if one analyzes them in the
connection with the Faddeev-Skyrme-Niemi effective model of the
low energy gluodynamics
\begin{equation}
L=\frac{1}{2}m^2 (\partial_{\mu} \vec{n})^2 -\frac{1}{4e^2}
[\vec{n} \cdot (\partial_{\mu} \vec{n} \times \partial_{\nu}
\vec{n})]^2. \label{fn}
\end{equation}
It is straightforward to see that all finite energy solutions in
this model must tend to a constant $\vec{n}_{\infty}$ at the
spatial infinity. Then field configurations being maps from $S^3$
into $S^2$ can be divided into disconnected classes and
characterized by the Hopf index. Indeed, many configurations with
a non-trivial topological charge, which appear to form knotted
structures have been numerically obtained \cite{battyde},
\cite{salo}.
\\
In order to reveal a close connection between the eikonal knots
and Faddeev-Niemi hopfions we rewrite the equations of motion for
model (\ref{fn}) in terms of the complex field $u$
(\ref{stereograf1}) \cite{aratyn}
\begin{equation}
(1+|u|^2) \partial^{\mu} L_{\mu} - 2u^* (L^{\mu} \partial_{\mu} u)
=0 \label{fn eq1}
\end{equation}
where
\begin{equation}
L_{\mu}=m^2\partial_{\mu}u -\frac{4}{e^2}
\frac{K_{\mu}}{(1+|u|^2)^2}$$ $$K_{\mu}=(\partial^{\nu} u
\partial_{\nu} u^*) \partial_{\mu} u - (\partial^{\nu} u )^2
\partial_{\mu} u^*.
\label{fn eq2}
\end{equation}
It has been recently observed \cite{aratyn} that such a model
possesses an integrable submodel if the following constrain is
satisfied
\begin{equation}
L_{\mu} \partial^{\mu} u=0. \label{int cond1}
\end{equation}
Then, infinite family of the local conserved currents can be
constructed \cite{alvarez}, \cite{ferreira}, \cite{sanchez1},
\cite{sanchez2}. On the other hand, it is a well-known fact from
the standard $(1+1)$ and $(2+1)$ soliton theory that the existence
of such a family of the currents usually leads to soliton
solutions with a non-trivial topology. Due to that one should
check whether also in the case of the Faddeev-Skyrme-Niemi model
the integrability condition can give us some hints how to
construct knotted solitons.
\\
For model (\ref{fn}) the integrability condition takes the form
\begin{equation}
m^2 (\partial_{\nu} u)^2=0.
\end{equation}
It vanishes if mass is equal to zero or the eikonal equation is
fulfilled. The first possibility is trivial since $m=0$ means the
absence of the kinetic term in Lagrangian (\ref{fn}) and no stable
soliton solutions can be obtained due to the instability under the
scale transformation. Thus, the eikonal equation appears to be the
unique, nontrivial integrability condition for the
Faddeev-Skyrme-Niemi model.
\\
However, it should be stressed that the full integrable submodel
consists of two equations. Namely, apart from the integrability
condition (\ref{int cond1}), also the dynamical equation has to be
taken into account
\begin{equation}
\partial_{\mu} \left[ m^2 \partial_{\mu}u -\frac{4}{e^2}
\frac{\partial^{\nu} u
\partial_{\nu} u^*}{(1+|u|^2)^2} \partial_{\mu}u \right]=0.
\label{fn dyn eq}
\end{equation}
The correct solutions of the Faddeev-Skyrme-Niemi model have to
satisfy both equations. Unfortunately, derived eik\-onal hopfions
do not solve the dynamical equation and in the consequence are not
solutions of the Faddeev-Skyr\-me-Niemi model. Nonetheless, the
fact that they appear in a very natural way in the context of the
Faddeev-Skyr\-me-Niemi model i.e. just as solutions of the
integrability condition might indicate a close relation between
them and the Faddeev-Niemi hopfions.
\\
This idea seems to be more realistic if we compare the eikonal
hopfions with the numerically found Faddeev-Nie\-mi hopfions
\cite{battyde}, \cite{salo}. It is striking that every hopfions
possesses its eikonal counterpart with the same topology and a
very similar shape.
\\
Moreover, there is an additional argument which strongly supports
the idea that the eikonal knots might applied in construction of
approximated Fadd\-eev-Nie\-mi hop\-fions. It follows from the
observation that the eikonal knots, if inserted into the total
energy integral calculated for the Faddeev-Skyrme-Niemi model,
give the finite value of this integral. The situation is even
better. The lowest energy eikonal configurations are only
approximately 20\% heavier than numerically derived hopfions. Let
us show it for $N=1$ Ansatz.
\\
Indeed, the Faddeev-Skyrme-Niemi model gives, for static
configurations, the following total energy integral
\begin{equation}
E= 2m^2 \int d^3x \frac{\nabla u \nabla u^*}{(1+|u|^2)^2}+
\frac{2}{e^2} \int d^3x \frac{(\nabla u \nabla u^*)^2 -(\nabla
u)^2 (\nabla u^*)^2}{(1+|u|^2)^4}, \label{energy fn1}
\end{equation}
where the stereographic projection (\ref{stereograf1}) has been
taken into account. Moreover, as our solutions fulfill the eikonal
equations
\begin{equation}
(\nabla u )^2=0, \label{eikonal new}
\end{equation}
thus
\begin{equation}
E= 2m^2 \int d^3x \frac{\nabla u \nabla
u^*}{(1+|u|^2)^2}+\frac{2}{e^2} \int d^3x \frac{(\nabla u \nabla
u^*)^2}{(1+|u|^2)^4}. \label{energy fn2}
\end{equation}
Now, we can take advantage of the eikonal hopfions and substitute
them into the total energy integral (\ref{energy fn2}). Let us
notice that
$$
\nabla u \nabla u^* =\frac{q^2}{\tilde{a}^2}\left[
f'^2+\left(m^2+\frac{k^2}{\sinh^2 \eta} \right) f^2 \right] =$$
\begin{equation}
= 2\frac{q^2}{\tilde{a}^2} \left(m^2+\frac{k^2}{\sinh^2 \eta}
\right) f^2 \label{kin1}
\end{equation}
and the Jacobian
\begin{equation}
d^3x =\frac{\tilde{a}^3 \sinh \eta}{q^3} d\xi d\phi d\eta.
\label{jakob}
\end{equation}
Then
\begin{equation}
E= 2m^2 \tilde{a} I_1+\frac{2}{\tilde{a} e^2} I_2, \label{energy3}
\end{equation}
where
\begin{equation}
I_1=2a^2 \int_0^{\infty} d\eta \int_0^{2\pi}\frac{d\xi}{q}
\int_0^{2\pi} d\phi
  \frac{\sinh \eta g^2
\left(m^2+\frac{k^2}{\sinh^2 \eta} \right)}{(1+c_0^2+a^2 g^2
+2ac_0g \cos[m\xi+k\phi])^2} \label{I1 a}
\end{equation}
and
\begin{equation}
I_2=4a^4 \int_0^{\infty} d\eta \int_0^{2\pi} d\xi q \int_0^{2\pi}
d\phi
 \frac{\sinh \eta g^4
\left(m^2+\frac{k^2}{\sinh^2 \eta} \right)^2}{(1+c_0^2+a^2 g^2
+2ac_0g\cos[m\xi+k\phi])^4}. \label{I2 a}
\end{equation}
As it was mentioned before we not only prove that the eikonal
knots provide finiteness of the total energy but additionally we
find the lowest energy configuration for fixed $k,m$. This
minimization procedure should be done with respect to three (in
the case of $N=1$) parameters: $\tilde{a}$ and $a,c_0$. At the
beginning we get rid of the scale parameter $\tilde{a}$
$$ \frac{\partial E}{\partial \tilde{a}}=0 \; \; \Rightarrow \; \;
\tilde{a}=\frac{1}{em} \sqrt{\frac{I_2}{I_1}}.$$ Then the total
energy takes the form
\begin{equation}
E=4\frac{m}{e} \sqrt{I_1I_2}. \label{energy4}
\end{equation}
Now, we calculate the previously defined integrals. It can be
carried out if one observes that
\begin{equation}
\int_0^{2\pi} \frac{d\phi}{(\alpha+\beta \cos[m\xi
+k\phi])^2}=\frac{2\pi \alpha}{(\alpha^2-\beta^2)^{\frac{3}{2}}},
\label{int 1}
\end{equation}
\begin{equation}
\int_0^{2\pi} \frac{d\phi}{(\alpha+\beta\cos[m\xi
+k\phi])^4}=\frac{\pi \alpha
(2\alpha^2+3\beta^2)}{(\alpha^2-\beta^2)^{\frac{7}{2}}},
\label{int 2}
\end{equation}
\begin{equation}
\int_0^{2\pi} \frac{d\xi}{\cosh \eta -\cos \xi} =
\frac{2\pi}{\sinh \eta}, \label{int 3}
\end{equation}
\begin{equation}
\int_0^{2\pi} d\xi (\cosh \eta -\cos \xi) = 2\pi \cosh \eta.
\label{int 4}
\end{equation}
Thus, one can finally obtain that
\begin{equation}
I_1=2a^2 (2\pi)^2 \int_0^{\infty} d \eta g^2
\left(m^2+\frac{k^2}{\sinh^2 \eta} \right)
\frac{1+c_0^2+a^2g^2}{\left[
(1+c_0^2+a^2g^2)^2-4c_0^2a^2g^2\right]^{\frac{3}{2}}} \label{I1 b}
\end{equation}
and
$$
I_2=4a^4 (2\pi)^2 \int_0^{\infty} d \eta \sinh \eta \cosh \eta g^4
\left(m^2+\frac{k^2}{\sinh^2 \eta} \right)^2 \; \times $$
\begin{equation}
\frac{(1+c_0^2+a^2g^2) [(1+c_0^2+a^2g^2)^2 +6c_0^2a^2g^2]}{\left[
(1+c_0^2+a^2g^2)^2-4c_0^2a^2g^2 \right]^{\frac{7}{2}}} \label{I2
b}
\end{equation}
\begin{table}
\center
\begin{tabular}{|c||c|c|c||c|}
  \hline
  type of a knot $(m,n)$& $a$ & $c_0$ & $E_{min}$&$E_{num}$\\
  \hline
  (1,1) & 1.252 & 0& 304.3 &252.0 \\  \hline
  (1,2) & 0.357 &0& 467.9  &417.5\\  \hline
  (2,1) & 5.23 & 0& 602.7 & 417.5\\  \hline
  (1,3) & 0.065 & 0& 658.1  &578.5\\  \hline
  (2,3) & 0.3 & 0& 1257.0  &990.5\\
  \hline
  \end{tabular}
  \caption{Minimal energy of the eikonal knots and Faddeev-Niemi
  hopfions}

\vspace*{0.3cm}

\begin{tabular}{|c||c|c|c|}
  \hline
  type of a knot $(m,n)$& $a$ & $c_0$ & $E$ \\
  \hline
  (1,1) & 1.252 &  0.2&311.2 \\  \hline
  (1,2) & 0.357 & 0.1&471.9 \\  \hline
  (2,1) & 5.23 &  0.2& 622.3\\  \hline
  (1,3) & 0.065 &  0.05&659.5 \\  \hline
  (2,3) & 0.3 &  0.1&1269.0 \\
  \hline
  \end{tabular}
  \caption{Energy of the knotted eikonal knots}
\end{table}
It may be easily checked that these two integrals are finite for
all possible profile functions of the eikonal hopfions.
\\
Now we are able to find the minimum of the total energy
(\ref{energy4}) as a function of $a,c_0$. It has been done by
means of numerical methods. The results $E_{min}$ for the simplest
knots are presented in Table 1 (we assume $m/e=1$). Let us shortly
comment obtained results.
\\
Firstly, we see that the eikonal knots are 'heavier' than knotted
solitons found in the numerical simulations $E_{num}$
\cite{battyde}. It is nothing surprising as the eikonal knots do
not fulfill the Faddeev-Skyrme-Niemi equations of motion that is
do not minimize the pertinent action. However, the difference is
small and is more less equal 20\% . Strictly speaking the accuracy
varies from 15\% for the lightest knots with $m=1$ up to 30-35\%
in the case of knots with bigger value of $Q_H$ or $m$ parameter.
This result is really unexpected since the eikonal knots are
solutions of such a very simple (first order and almost linear)
equation.
\\
Secondly, the lowest energy configurations are achieved for
$c_0=0$. As we know it means that a knot is located at
$\eta_0=\infty$. In other words, for fixed $m,k$ the unknot (i.e.
configurations where surfaces $n^3=const.$ are toruses) possesses
lower energy than other knotted eikonal solutions. It is a little
bit discouraging since the Faddeev-Niemi hopfions are in general
really knotted solitons. However, one can observe that even very
small increase of the energy $E$ causes that $c_0 \neq 0$ (see
Table 2). Then, what is more important, also the ratio
$\frac{c_0}{a}$ differs from zero significantly. It guarantees
that the knotted structure of a eikonal solution becomes restored.
\\
Thirdly, there is no $m \leftrightarrow k$ degeneracy. The eikonal
knots with $m=p, k=q$ and $m=q,k=p$ do not lead to the same total
energy. In particular, for configurations with the fixed
topological charge, the lowest energy state is a knot with $m=1$.
In the case of knots with bigger value of the parameter $m$ the
total energy grows significantly.
\\
We see that the eikonal knots seem to be quite promising and can
be applied to the Faddeev-Skyrme-Niemi model. Since our solutions
possess a well defined topological charge and approximate the
shape as well as the total energy of the Faddeev-Niemi hopfions
with on an average 20\% accuracy, one could regard them as first
step in construction of approximated solutions (given by the
analytical expression) to the Faddeev-Skyrme-Niemi model.
\\
Recently Ward \cite{ward2} has analyzed the instanton
approximations to Faddeev-Niemi hopfions with $Q_H=1,2$ Hopf
index. It would be very interesting to relate it with the eikonal
approximation.
\\
It should be noticed that similar construction provides
approximated, analytical solutions in non-exactly solvable $(2+1)$
dimensional systems. In fact, the Baby Skyrme model
\cite{baby_skyrme} and Skyrme model in $(3+1)$ dimensions
\cite{skyrme}, \cite{manton} can serve as very good examples.
%%%%%%%%%%%%%%%%%%%%%%%%%%%%%%%%%%%%%%%%%%%%%%%%%%%%%%%%%%%%%%%
\section{\bf{Conclusions}}
%%%%%%%%%%%%%%%%%%%%%%%%%%%%%%%%%%%%%%%%%%%%%%%%%%%%%%%%%%%%%%%
In this work multi-soliton and multi-knot configurations,
generated by the eikonal equation in two and three space
dimensions respectively, have been discussed. It has been proved
that various topologically non-trivial configurations can be
systematically and analytically derived from the eikonal equation.
\\
In the model with the two space dimensions (which is treated here
just as a toy model for the later investigations) multi-soliton
solutions corresponding to the $O(3)$ sigma model have been
obtained. In particular, we took under consideration one and two
component Ansatz i.e. $N=1,2$. In this case we restored the
standard result that the energy of a multi-soliton solution
depends only on the total topological number. The way how the
topological charge is distributed on the individual solitons does
not play any role. Thus, for example, the energy of a single
soliton with winding number $n$ is equal to the energy of a
collection of $n$ solitons with the unit charge. In both cases we
observe saturation of the energy-charge inequality. Moreover,
energy remains constant under any changes of positions of the
solitons. It is exactly as in the Bogomolny limit where
topological solitons do not attract or repel each other. Due to
that the whole moduli space has been found.
\\
In the most important, three dimensional space case we have found
that the eikonal equation generates multi-knot configurations with
an arbitrary value of the Hopf index. As previously, Ansatz
(\ref{anzatz}) with one and two components has been investigated
in details. Let us summarize obtained results.
\\
Using the simplest, one component Ansatz (\ref{anzatz}) we are
able to construct one as well as multi-knot configurations which,
in general, consist of the same (topologically) knots linked
together. The elementary knot can have various topology. For
example a trefoil knot has been derived. It is unlikely the
standard analytical hopfion solutions which have always toroidal
symmetry and are not able to describe such a trefoil state.
\\
By means of the two component Ansatz multi-knot configurations
with a central knot located at $\eta=\infty$ and a few satellite
knots winded on a torus $\eta=const$ have been obtained. In the
contrary to the central knot, which is always circle, satellite
solitons can take various, topologically inequivalent shapes known
from the $N=1$ case.
\\
In addition, we have argued that the multi-knot solutions can be
useful in the context of the Faddeev-Skyrme-Niemi model. Thus,
they appear to be interesting not only from the mathematical point
of view (as analytical knots) but might also have practical
applications. We have shown that the eikonal knots provide an
analytical framework in which the qualitative features of the
Faddeev-Niemi hopfions can be captured. Moreover, also
quantitative aspects i.e. the energy of the hopfion can be
investigated as well. Although the eikonal knots are approximately
20\% heavier than the numerical hopfions, what is rather a poor
accuracy in compare with the rational Ansatz for Skyrmions, one
can expect that for other shape functions a better approximation
might be obtained.
\\
There are several directions in which the present work can be
continued. First of all one should try to achieve better
approximation to the knotted solutions of the Faddeev-Skyrme-Niemi
effective model. It means that new, more accurate shape functions
have to be checked. It is in accordance with the observation that
presented here knots solve only the integrability condition
(eikonal equation) but not the pertinent dynamical equations of
motion. Due to that it is not surprising that the eikonal shape
function is an origin for some problems (eikonal knots are too
heavy and tend to unknotted configurations). One can expect that
these new shape functions will not only better approximate the
energy of the Faddeev-Niemi hopfions but also guarantee non-zero
value of the parameter $c_0$ and ensure the knotted structure of
the solutions. We would like to address this issue in our next
paper.
\\
There is also a very interesting question concerning the shape of
the eikonal hopfions. Cores of all presented here knots are
situated on a torus with a constant radius. However, there are
many knots which cannot be plotted as a closed curve on a torus.
Thus one could ask whether it is possible to construct such knots
(non-torus knots) in the framework of the eikonal equation.
\\
Of course, one might also apply the eikonal equation to face more
advanced problems in the Faddeev-Skyrme-Niemi theory and
investigate time-depen\-dent configurations as for instance
scattering solutions or breather.
\\
On the other hand, one can try to find a Lagrangian which
possesses obtained here topological configurations as solutions of
the corresponding field equations. One can for example consider
recently proposed modifications of the Faddeev-Skyrme-Niemi model
which break the global $O(3)$ symmetry \cite{niemi3}, \cite{wipf},
\cite{my}. Application of the eikonal equation to other models of
glueballs \cite{dzhu}, \cite{bazeia}, based in general on the
$\vec{n}$ field, would be also interesting.
\\ \\
I would like to thank Prof. A. Niemi for discussion. I am also
indebted to Dr. C. Adam and Prof. H. Arod\'{z} for many valuable
and helpful remarks.
\\
This work is partially supported by Foundation for Polish Science
FNP and ESF "COSLAB" programme.

\end{document}